\documentclass[aps,prd,superscriptaddress,showpacs,amsmath,amssymb]{revtex4}
\pdfoutput=1
\usepackage{epsfig}
\usepackage{color}
\usepackage{latexsym}
\usepackage{amsmath}
\usepackage{amsfonts}
\usepackage{amssymb}
\usepackage{graphicx}
\newcommand{\be}{\begin{eqnarray}}
\newcommand{\ee}{\end{eqnarray}}
\newcommand{\ud}{\mathrm{d}}

\newcommand{\pd}{\partial}
\begin{document}

\title{Stability of Einstein-Maxwell-Kalb-Ramond Wormholes}
\author{Paul H. Cox}
\email{phcox@tamuk.edu}
\affiliation{Department of Physics and Geosciences, Texas A\&M University-Kingsville, M.S.C. 175, Kingsville, TX 78363-8202, USA
}
\author{Benjamin C. Harms}
\email{bharms@ua.edu}
\affiliation{Department of Physics and Astronomy, The University
of Alabama, Box 870324, Tuscaloosa, AL 35487-0324, USA}
\author{Shaoqi Hou}
\email{shou@crimson.ua.edu}
\affiliation{Department of Physics and Astronomy, The University
of Alabama, Box 870324, Tuscaloosa, AL 35487-0324, USA}
\begin{abstract}
This paper investigates a particular type of wormhole.  The wormhole solutions studied are obtained by 
sewing together two static, 
spherically symmeric, charged black-hole metrics 
at their horizons.   The charged wormholes are in a background Kalb-Ramond field, which is the source of the necessary tension in the gravitational field.  The 
metric-tensor 
elements are studied by numerically solving Einstein's equations with stress-energy-tensor 
elements given by the combination of static electric and Kalb-Ramond fields.   For a certain range of electric charge the tension is positive away from the wormhole throat, but the tension is negative near the throat, making it non-traversable. The wormholes are found to be quasi-stable against decay via gravitational instanton tunnelling.
\end{abstract}
\pacs{04.20.Jb, 04.40.Nr,04.60.Bc}
\maketitle
%
\section{Introduction}
Wormhole solutions of the Einstein equations were first studied by Einstein and Rosen \cite{ein}, who called such solutions `bridges'.  The solutions in \cite{ein} were offered as models for the elementary particles known at the time (the proton and the electron).  Wormhole models of neutral and charged particles were also studied in \cite{arn1,arn2}, where the relation between the mass and the charge which is characteristic of such models was first obtained \cite{casadio}. These and subsequent attempts to identify elementary particles as wormholes have failed.  Since wormholes can be constructed which join two separate space-times or places in a space-time, in a quantum theory of gravity they allow the formation of `baby universes' \cite{hawk}.  Wormholes have also been proposed as a means of interstellar travel \cite{morris}, although such objects would require some form of `exotic' matter in order to allow interstellar travellers to safely pass through.  The 
wormhole solutions
studied in this paper are obtained by sewing together two `charged' black hole metrics at their horizons.  The resulting space-time consists of two congruent regions which are geometrically connected but are causally disconnected. The wormholes are assumed to be in a background field with tension, which is provided by a massless, antisymmetric-tensor field -- the 
Kalb-Ramond (K-R) 
field.  Such fields occur naturally in string theory and should be a constituent of a gravitational multiplet if string theory is a viable theory of all physical phenomena.  
\label{intro}
\section{Charged Wormholes}
\label{charge}
\subsection{Metric Tensor Elements}
Morris and Thorne \cite{morris} investigated traversable wormholes, finding that they require 'exotic' matter, with negative energy density (at least as seen by some observers) as well as significant levels of tension (negative pressure).  Some types of fields allow tension, and Rahaman, Kalam, and Ghosh \cite{rahaman} showed that certain static, spherically symmetric solutions of the K-R field \cite{kalb} are wormholes (though they are not traversable).  We study such (non-traversable) wormholes with a metric of the spherically symmetric form \cite{morris} (with $c = 1$; we also frequently use $G = 1$)
\be
ds^2 = -{\rm e}^{A(r)}\,dt^2 +  {\rm e}^{B(r)}\,dr^2 +\,r^2\,\left(d\theta^2 + \sin^2\theta\,d\phi^2\right)\, ,
\label{mt}  
\ee
The functions $A(r)$ and $B(r)$ are determined by the Einstein equations (for brevity we use $'$ for $d/dr$) 
\be
&\displaystyle{\frac {{{\rm e}^{A \left( r \right) }} B' \left( r \right) }{r{{\rm e}^{B \left( r \right) }}}}+{\frac {{{\rm e}^{A
 \left( r \right) }}}{{r}^{2}}}-{\frac {{{\rm e}^{A \left( r \right) }}}{{r}^{2}{{\rm e}^{B \left( r \right) }}}}
\,=\,8\pi T_{00}\nonumber\\
&\displaystyle{\frac {A' \left( r \right) }{r}}-{\frac {{{\rm e}^{B \left( r \right) }}}{{r}^{2}}}+\frac{1}{r^{2}}
\,=\,8\pi T_{11}\label{eeqns}\\
&\displaystyle-\frac{1}{2}\,{\frac { r B' \left( r \right) }{{{\rm e}^{B \left( r \right) }}}}+\frac{1}{2}\,{\frac { 
r A' \left( r \right)  }{{{\rm e}^{B \left( r \right) }}}}+\frac{1}{2}\,{
\frac {{r}^{2} A'' \left( r \right) }{{{\rm e}^{B \left( r \right) }}}}+\frac{1}{4}\,{\frac {{r}^{2} \left( A'
 \left( r \right)  \right) ^{2}}{{{\rm e}^{B \left( r \right) }}}}-\frac{1}{4}\,{\frac {{r}^{2} \left( A' \left( r \right)  \right) 
B' \left( r \right) }{{{\rm e}^{B \left( r \right) }}}}
\nonumber\\
&\,=\,8\pi T_{22}\, =\, 8\pi T_{33}/\sin^2(\theta) \, .\nonumber
 \ee
 The action for such a wormhole 
in background electromagnetic and K-R fields is given by
\be 
S =\int_{\mathcal{M}}\sqrt{-g}\left( \frac{R}{16\,\pi\,G} -\frac{1}{4}F_{\mu\nu}F^{\mu\nu}-\frac{1}{12}\,H_{\mu\nu\lambda}\,H^{\mu\nu\lambda}\right)\,d^4x \, ,
\label{action}
\ee
where $g$ is the metric tensor determinant, $R$ is the Ricci scalar, $F_{\mu\nu}$ is the 
electromagnetic-field 
tensor, and $H_{\mu\nu\lambda}$ is the 
totally antisymmetric K-R-field tensor. 
 The energy-momentum tensor elements are given by 
\be 
T_{\mu\nu}\,=\,F_{\mu\lambda}\,{F_{\nu}}^{\lambda}-\frac{1}{4}g_{\mu\nu}\,F_{\lambda\sigma}F^{\lambda\sigma}+\frac{1}{12}\left( 3\,H_{\lambda\sigma\mu}{H^{\lambda\sigma}}_{\nu}-\frac{1}{2}\,g_{\mu\nu}H_{\lambda\sigma\tau}H^{\lambda\sigma\tau}\right) \, .
\ee
The electromagnetic and K-R fields satisfy the equations
\be 
&\frac{1}{\sqrt{-g}}\partial_{\nu}\left(\sqrt{-g}\,F^{\mu\nu}\right)\,=\,0 \, , \nonumber\\
&\frac{1}{\sqrt{-g}}\partial_{\nu}\left(\sqrt{-g}\,H^{\mu\lambda\nu}\right)\,=\,0 .
\label{FE}
\ee

Many relationships among local properties look more familiar in a local orthonormal coordinate basis, the proper reference frame of observers at fixed $r, \theta, \phi$:
\be 
 e_{\hat{t}} &=& {\rm e}^{-A(r)/2}\, e_t\: , \:\: e_{\hat{r}} = {\rm e}^{-B(r)/2}\, e_r\nonumber \\
\qquad
 e_{\hat{\theta}} &=& r^{-1}\, e_{\theta}\: , \hspace{7mm} e_{\hat{\phi}} = (r\,\sin(\theta))^{-1}\, e_{\phi} \, .
\ee
In such a basis, pressure $p$ is given by the equal $\hat{\theta}\hat{\theta}$ and $\hat{\phi}\hat{\phi}$ components of the stress-energy tensor, while energy density $\rho$ and tension $\tau$ are given by the  $\hat{t}\hat{t}$ and negative $\hat{r}\hat{r}$ components of the Einstein tensor respectively.
We choose to consider the static field case in which the only independent non-zero components of $F^{\hat{\mu}\hat{\nu}}$ and $H^{\hat{\mu}\hat{\nu}\hat{\lambda}}$ are $F^{\hat{0}\hat{1}}\,=\,E(r)\,=Q/2\sqrt\pi r^2\, , \; H^{\hat{0}\hat{2}\hat{3}}\,=\,h(r)/(\sin(\theta))$.
The energy-momentum tensor elements are then given by
\be 
T_{00}&=&{{\rm e}^{A \left( r \right) }} \left( \, \left( E \left( r
 \right)  \right) ^{2}/2+\, \left( h \left( r \right)  \right) ^{2}/2
 \right)
 = {{\rm e}^{A \left( r \right) }} T_{\hat{0}\hat{0}} 
 \nonumber \\
T_{11}&=&{{\rm e}^{B \left( r \right) }} \left( -\, \left( E \left( r
 \right)  \right) ^{2}/2+\, \left( h \left( r \right)  \right) ^{2}/2
 \right) 
 = {{\rm e}^{B \left( r \right) }} T_{\hat{1}\hat{1}}
 \label{emten}\\
T_{22}&=&{r}^{2} \left( \, \left( E \left( r \right)  \right) ^{2}/2-\,
 \left( h \left( r \right)  \right) ^{2}/2 \right)
 = {r}^{2}  T_{\hat{2}\hat{2}} = {r}^{2}  T_{\hat{3}\hat{3}}
 \, . \nonumber
\ee

The Einstein equations (\ref{eeqns}) cannot be solved analytically; however, the numerical solution of these equations shows that there is a singularity in the function $B(r)$ at a point $r_0$ determined in part by the value of the charge $Q$.  This singularity at $r = r_0$ is handled by sewing together two congruent copies of the region $r \geq r_0$ (Fig.1).  (Continuity at this junction can be seen by embedding the manifold in a five-dimensional spatially flat space-time with an additional coordinate $z$ joining $t, r, \theta, \phi$; the embedding $z = z(r)$ is defined by  \cite{morris}
\be 
\frac{dz}{dr} = \pm\,\left(\sqrt{b(r)-1}\right)\, ,
\qquad z(r_0) = 0 ,
\label{dzdr}
\ee
where $b(r)\,=\,{\rm e}^{B(r)}$; the +/- sign distinguishes the two copies\cite{rahaman}.  
\begin{figure}[ht]
\centering
\includegraphics[viewport=0cm 0cm 18.59cm 18.34cm,scale=0.55,clip]{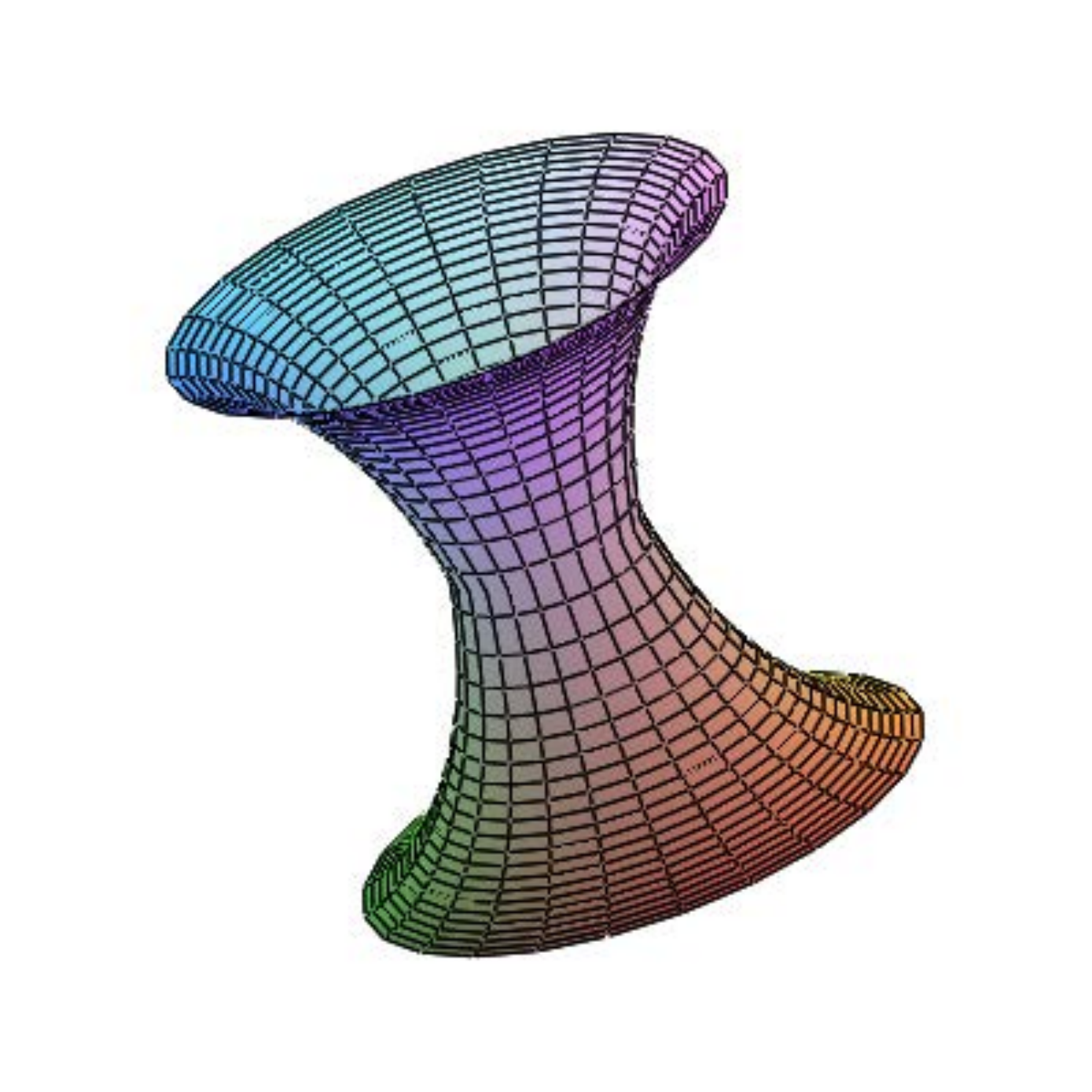}
\caption{The throat region of a wormhole obtained by sewing together two  black holes at their horizons.}
\label{wh}
\end{figure}
The numerical solution of the Einstein gravitational field equations allows the functions $A(r)$ and $b(r)$ defining the $g_{00}$ and $g_{11}$ metric tensor elements, respectively,  to be plotted versus the radius $r$ as shown in Figs. 2 and 3.    
\begin{figure}[ht]
\centering
\includegraphics[viewport=0cm 0cm 15.59cm 20.34cm,scale=0.55,clip]{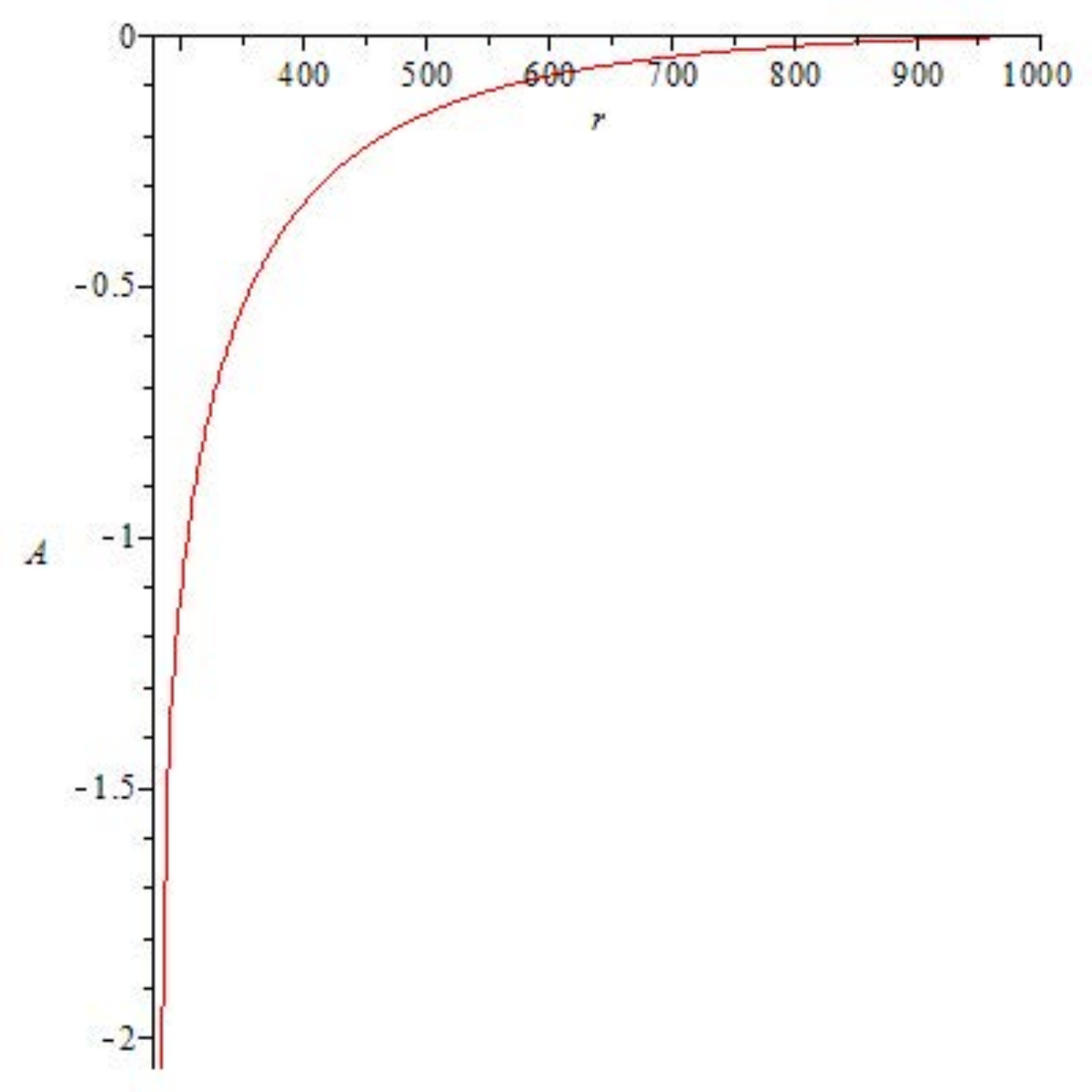}
\caption{The function 
$A(r)\,=\,\ln(-g_{00})$ 
plotted versus the radial coordinate $r$ in dimensionless units.}
\label{Ar}
\end{figure}
\begin{figure}[ht]
\centering
\includegraphics[viewport=0cm 0cm 15.59cm 20.34cm,scale=0.55,clip]{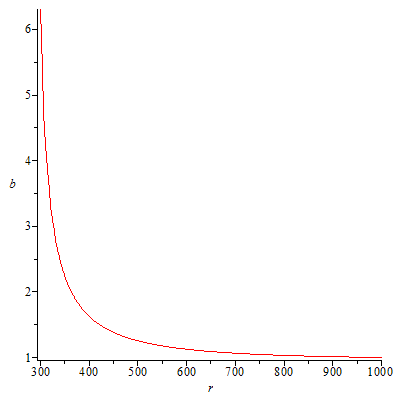}
\caption{The function $b(r)\,=\,g_{11}\,=\,{\rm e}^{B(r)}$ plotted versus the radial coordinate $r$ in dimensionless units.}
\label{br}
\end{figure}
The tension $\tau(r)$ is positive for $r\,>\,r_0$ if the electric charge $Q$  is large enough (Fig. 4), but $\tau(r)$ is always less than the energy density $\rho(r)$ as can be seen from 
Eq.~(\ref{emten}), so the wormhole described by this model is not traversable.  
\begin{figure}[ht]
\centering
\includegraphics[viewport=0cm 0cm 15.59cm 20.34cm,scale=0.55,clip]{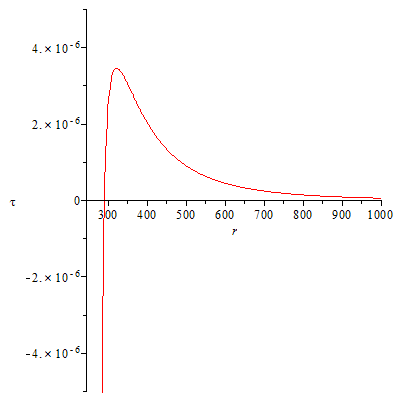}
\caption{The tension $\tau(r)$ plotted versus the radial coordinate $r$ in dimensionless units.}
\label{taur}
\end{figure}
The total electric and K-R charges for the space-time described by Eq.~(\ref{mt}) are zero.  However, a distant observer would be on either one side of the causal boundary (at $r_0$) or the other and would therefore detect a net electric charge of one sign or the other for each of the wormhole termini.  The K-R `charge' of the wormhole is zero due to the fact that the only non-zero component of $H^{\mu\nu\lambda}$ is transverse to the radial coordinate.

\subsection{Einstein-Maxwell-Kalb-Ramond Action}

The metric in Eq.~(\ref{mt}) is part of the solutions of the field equations obtained from the Einstein-Maxwell-Kalb-Ramond action (Eq.~(\ref{action})) on a manifold $\mathcal{M}$. We have assumed a static solution with spherical symmetry, which leads to fields with only one independent non-zero component each, $F^{01}$ and $H^{023}$ for the electromagnetic and K-R fields respectively. These fields are normalized as
\be 
F_{\mu\nu}\,F^{\mu\nu}&=& -2\, E(r)^2\nonumber \\
H_{\mu\nu\lambda}\,H^{\mu\nu\lambda}&=&- 6\, h(r)^2\, .
\label{FH}
\ee
$F_{\mu\nu}$ and $H_{\mu\nu\lambda}$ satisfy the field equations given in Eq.~(\ref{FE}).

The action is calculated using the procedure of \cite{gib}. For this, the form of the action is required to be first order in field derivatives \cite{gib}; this requires an integration by parts, leading to a non-trivial boundary term, 
\be 
S_B = \frac{1}{8\,\pi\,G}\int_{\partial\mathcal{M}}\,\sqrt{-\gamma}\,K\,d^3x
\label{bdA}
\ee
where $\gamma$ is the induced metric on the boundary $\partial\,\mathcal{M}$ and $K$ is the trace of the second fundamental form, 
which is
\be 
K_{\mu\nu} = \nabla_{\mu}n_\nu -n_\lambda\,n^\lambda\,n_\mu\,n^\sigma\,\nabla_\sigma\,n_\nu \, .
\label{sec}
\ee
Here, vector $n_\mu$ is a unit vector normal to the boundary $\partial\,\mathcal{M}$; in the present case $n_\mu = (0,{\rm e}^{B(r)/2},0,0)$.  The action in Eq.~(\ref{bdA}) diverges when the radial coordinate becomes large, so it is renormalized by subtracting the flat-space value of the trace of the second fundamental form.  Eq.~(\ref{bdA}) then becomes
\be 
S_B = \frac{1}{8\,\pi\,G}\int_{\partial\,M}\,\sqrt{-\gamma}\,[K]\,d^3x
\label{bdAn}
\ee
where $[K]= K-K^F|_{r\to\infty}$ and $K^F$ is the second fundamental form for flat space. 

The renormalized action is then
\be 
S=\int_\mathcal{M}\ud^4x\sqrt{-g}\Big[\frac{R}{16\pi}-\frac{1}{4}F_{\mu\nu}F^{\mu\nu}-\frac{1}{12} H_{\rho\sigma\lambda} H^{\rho\sigma\lambda}\Big]+\frac{1}{8\pi}\int_{\pd\mathcal{M}}[K]\ud\Sigma \, ,
\ee 
where $\Sigma$ is the volume element of the compact Euclidean spatial hypersurface for a section with imaginary time $\tau$ and at fixed $r$  \cite{gib}
\be 
\ud\Sigma=-ie^{A/2}r^2\sin\theta\,\ud \tau\,\ud\theta\,\ud\phi \, .
\ee
The scalar curvature $R$ equals 
\be 
R=-\frac{2\pi}{3}  H_{\rho\sigma\lambda} H^{\rho\sigma\lambda}\, ,
\ee
giving a total action of
\be 
S = \int_\mathcal{M}\ud^4x\sqrt{-g}\Big[-\frac{1}{4}F_{\mu\nu}F^{\mu\nu}-\frac{1}{4} H_{\rho\sigma\lambda} H^{\rho\sigma\lambda}\Big]+\frac{1}{8\pi}\int_{\pd\mathcal{M}}[K]\ud\Sigma \, .
\ee
In terms of the functions defined in 
Eq.~(\ref{FH}) 
the action is 
\be 
S = \int_\mathcal{M}\ud^4x\sqrt{-g}\Big(E(r)^2+6\pi h^2\Big)+\frac{1}{8\pi}\int_{\pd\mathcal{M}}[K]\ud\Sigma \, .
\label{na}
\ee

The function $h(r)$ is determined from the properties that the K-R field can be expressed as the dual of the derivative of a pseudoscalar $\Phi(r)$
\be 
H_{\mu\nu\lambda}\,=\,{\epsilon_{\mu\nu\lambda}}^{\rho}\,\pd_{\rho}\Phi(r)
\ee
and that $H_{\mu\nu\lambda}$ satisfies the Bianchi identity
\be 
\epsilon^{\mu\nu\lambda\alpha}\,\pd_{\alpha}H_{\mu\nu\lambda}\,=\,0 \, .
\ee
These relations require that 
\be 
h(r)\,=\frac{\,h_0\,{\rm e}^{-A(r)/2}}{r^2}\, ,
\label{hr}
\ee
where $h_0$ is a constant.  Substituting the above expressions for $E(r)$ and $h(r)$ into Eq.~(\ref{na}) and evaluating the surface term, the expression for the action becomes
\be 
S\,=\,i\frac{8\pi^2}{\kappa}\int_{r_0}^\infty e^{(A+B)/2}r^2\Big(\frac{Q^2}{2r^4}+12\pi \frac{\,h_0^2\,{\rm e}^{-A(r)}}{r^4}\Big)\ud r-i\frac{\pi}{\kappa}e^{A/2}r\Big[2(e^{-B/2}-1)-\frac{rA'}{2}e^{-B/2}\Big]\, \, ,
\label{Actr} 
\ee
where $\kappa$ is the surface gravity evaluated at $r\,=\,r_0$
\be 
\kappa=\left[\Big|\frac{A'}{2}\Big|e^{(A-B)/2}\right]\Big|_{r_0}\, .
\ee

The functions $A(r)$ and $B(r)$ are determined by the Einstein equations and are given by
\be 
e^{-B(r)}&=&1+c_1/r+\eta(r)/r-Q^2/8\pi r^2\nonumber\\
e^{A(r)}&=&\frac{c_2\,e^{B(r)}}{r^2}\exp\Big[\int(2+Q^2/4\pi r^2)\frac{e^{B(r)}}{r}\ud r\Big]\, ,
\ee
where 
$c_1$ and $c_2$ are constants of integration and the function $\eta(r)$ is defined by 
\be 
\eta(r)\,=\,4\pi\int r^2 \,h(r)^2\ud r\, .
\label{eta}
\ee
This function satisfies the differential equation 
\be 
\Big[2\,r\,(r+c_1+\eta)-\frac{Q^2}{4\pi}\Big]\,\eta''+2\tau'\,(r+c_1+\eta)-2\,r\,\eta'\,(\eta'-1)=0\, ,
\ee
and can be expressed as an infinite series
\be 
\eta(r)=\sum_{m=1}^\infty\frac{b_m}{r^m}\, ,
\ee
whose expansion coefficients $b_m$ satisfy the recursion relation
\be 
2\,m(m-1)b_m+2(m-1)^2\,b_{m-1}\,c_1
&-& \frac{Q^2}{4\pi}\,(m-1)\,(m-2)\,b_{m-2}
\nonumber \\  &-&
2\sum_{n=1}^{m-2}(m-2n-1)\,(n+1)\,b_{m-n-1}\,b_n=0\, ,\: m\ge2 \, .
\ee
For $c_1\,=\,0$ an approximate solution can be obtained for $\eta(r)$ for small $Q$ by making the change of variable 
\be 
\eta(r)\,=\,-b/r+(Q^2/4\pi)\zeta(r), 
\ee
where $b$ is a wormhole parameter; for $Q = 0$ it is the throat radius. 
The equation for $\zeta(r)$ is
\be 
(r^2-b)\,\zeta''+(2\,r-3\,b/r)\zeta'+(Q^2/4\pi)(r\,\zeta\,\zeta''-r\,\zeta'^2+\zeta\,\zeta')-(b/r^2)\zeta=0\, .
\label{zeta}
\ee
When in this equation $Q^2$ is neglected as being much smaller than $b$,  the solution to Eq.~(\ref{zeta}) is
\be 
\zeta(r)=\frac{a_1}{r}+\frac{a_2}{r}\Big[\frac{1}{\chi(r)}+\arctan\chi(r)\Big] \, ,
\ee
where $a_1$ and $a_2$ are constants of integration and $\chi(r)\,=\,\sqrt{b/(r^2-b)}$.  In the expression for $\zeta(r)$ the constant $a_1$ can be set to zero without loss of generality, or, equivalently, it can be absorbed into the constant $b$.  With this approximate expression for $\zeta(r)$ the terms appearing in the action can be written as (to first order in $Q^2$)
\be 
h(r)^2&\approx&\frac{4}{\kappa r^4}\Big[b-(Q^2/4\pi)\Big(a_2\arctan\chi\Big)\Big] \, ,
\label{hr2}
\\
e^B(r) &\approx& 1+\chi^2+\frac{Q^2}{4\pi b}\chi^2(1+\chi^2)\Big[1-2a_2\Big(\frac{1}{\chi}+\arctan\chi\Big)\Big] \, ,
\\
e^A(r) &\approx& 1+\frac{Q^2}{8\pi b}\Big[2a_2\arctan\chi-\ln(1+\chi^2)\Big] \, ,
\\
r_0&\approx&\sqrt{b}+\frac{Q^2}{16\pi\sqrt{b}}(1-\pi a_2),
\label{rp}
\\
\kappa &\approx& \left[\frac{Q^2}{8\pi b^{3/2}}\frac{\chi^2}{1+\chi^2}|\chi-a_2|\right]\Big|_{r_0},
\ee

Inserting these expressions into Eq.~(\ref{Actr}) and carrying out the integration over $r$ gives
\be 
S\,=\,&&\Bigg\{\Big(\frac{Q^2}{8\pi\sqrt b}+\frac{36\sqrt{b}}{\kappa}\Big)\arctan\chi+\frac{9\,Q^2}{4\,\pi\kappa \sqrt b}\Big[\chi-\arctan\chi-2a_2\chi\arctan\chi\Big]\Bigg\}_0^{\chi_{r_0}}
\nonumber\\
&&-\frac{9\,Q^2}{4\,\pi\kappa \sqrt b}\int_0^{\chi_{r_0}}\frac{\ln(1+\chi^2)}{1+\chi^2}\ud\chi \, .
\label{Actp}
\ee
At $r_0$ the expressions for $\chi(r)$ and $\kappa$ become
\be 
\chi_0\approx\sqrt{\frac{8\pi b}{Q^2(1-\pi a_2)}}\gg 1 \, ,
\ee
and 
\be 
\kappa\approx \frac{|Q|}{2b\sqrt{2\pi(1-\pi a_2)}} \, \label{kappa} .
\ee
With these expressions the action in Eq.~(\ref{Actp}) is approximately 
\be 
S\approx i\frac{8\pi}{\kappa}\Bigg\{\frac{\pi}{8\pi}\Big(\frac{Q^2}{2\sqrt b}+\frac{36\sqrt{b}}{\kappa}\Big)-\frac{9\,Q^2}{4\pi\kappa b}\Big[\pi\ln2+\frac{\pi}{2}-(1-\pi a_2)\frac{2b^{3/2}}{Q^2}\kappa\Big]\Bigg\}-i\frac{\pi}{\kappa}\frac{a_2Q^2}{4\pi\sqrt b}\, .
\label{Acta}
\ee

The constants $a_2$ and $b$ can be related to the ADM mass of the wormhole and to the constant $h_0$.  The ADM mass is given byan integral over the $r \rightarrow \infty$ boundary $B$ of the manifold, 
\be 
M\,=\,\oint_B r^i(\pd_jh_i^j-\pd_ih^j_j)=-8\pi(c_1+\eta)|_{r\rightarrow\infty}\,=\,-2Q^2 a_2/\sqrt{b}\, .
\label{mass}
\ee
where $h_{ij}$'s are the induced metric tensor elements on the boundary $B$, which, in an asymptotic Cartesian system, are given by $h^{ij}=-\frac{c_1+\eta}{r^3}x^ix^j$.  An approximate expression for $h_0$ can be obtained from Eqs.~(\ref{hr}) and (\ref{eta})
\be 
h_0^2 \approx \frac{b}{24\,\pi} \, .
\ee
The K-R `charge' is zero for this wormhole, because the only non-zero component of the field is transverse to the radial direction.  The K-R field in the wormhole must comefrom an external source.
  
\subsection{Wormhole Stability}
The stability of a wormhole can be analyzed within the context of quantum gravity by evaluating the partition function obtained from the summation over all histories
\be 
Z = \int[dg]\,\exp(i\,S[g])
\label{part}
\ee
in units where $\hbar\, =\,c\,=\,1$. 
The decay probability of a body is determined by the probability of a gravitational instanton \cite{gibb} tunnelling from it to the vacuum.  This probability can be written as \cite{jack,callan}
\be 
P \sim {\rm e}^{i\,S_E/\hbar}
\label{prob}
\ee
where $S_E$ is the Euclidean action, which for our EMKR wormholes is given by Eq.~(\ref{Acta}). After substituting in for $\kappa$ (Eq.~(\ref{kappa})) 
and $a_2$ (Eq.~(\ref{mass})), the dominant term in the 
action (\ref{Acta}) is
\be 
S_E\,\approx\,i  \frac {144\,\sqrt{2}{\pi }^{3/2}\,c^7{b}^{2} M}{G_N^{3/2}\,{Q}^{3}}\, .
\ee
where the fundamental constants $c$ and $G_N$ are now explicitly displayed. This expression shows that wormholes produced with small charge and any measurable mass and geometric size are quasi-stable, since the probability for decay is extremely small.   The possibility of forming quasi-stable, microscopic wormholes is due to the presence of the K-R field with parameter  $h_0\,=\,\sqrt{b/24\pi}$.  The horizon radius, $r_0$, is determined, at least in the approximation used in  Eq.~(\ref{rp}),  principally by the K-R field strength, $h_0$, and the mass, $M$. Because of the presence of the K-R field, the charge to mass ratio no longer has to be $< 1$, unlike the case for charged black holes with no K-R field.   

\section{Conclusions}
Since wormholes may exist or perhaps can be created in nature or artificially as particles, as baby universes, or as mechanisms for time travel, an analysis of their stability is essential. We have shown in the above analysis that a particular type of wormhole can 
be quasi-stable through achieving a suitable 
adjustment of the parameters of the system.  The parameter $h_0$ 
associated with the K-R field, in the approximation that $Q^2 \ll b$, essentially determines the radius of the wormhole.   Moderate values for the K-R coupling 
strength, $h_0^2\approx 10^{-18}$ erg cm, correspond 
to $r_0 \approx l_p$, where $l_p$ is the Planck length, and to energy densities a few orders of magnitude less than that for a Planck mass. The mass of the wormhole is not determined by the K-R coupling strength and so could be anything between zero and the Planck mass.   Wormholes with masses on the order of standard model masses ($ < 1$ TeV) could be created by high energy cosmic rays or in an accelerator,  providing a realization of the original attempt of Einstein and Rosen  \cite{ein} to describe a particle as a wormhole, needing only the requirement that a K-R field of suitable strength is present. 

%
%

%
%
\acknowledgments
We thank Allen Stern and Roberto Casadio for discussions on the work described in this paper.  This work was supported in part by the U.S. Department
of Energy under Grant No. DE-FG02-10ER41714 .

\end{document}